\documentclass{article}

\usepackage[english]{babel}
\usepackage{xcolor}
\usepackage[letterpaper,top=2cm,bottom=2cm,left=3cm,right=3cm,marginparwidth=1.75cm]{geometry}
\usepackage{verbatim}
\usepackage{amsmath}
\usepackage{graphicx}
\usepackage[colorlinks=true, allcolors=blue]{hyperref}
\usepackage{subcaption}
\usepackage{cancel}
\usepackage[final]{pdfpages}
\usepackage{authblk}
\usepackage{csquotes}
\usepackage{xcolor}
\usepackage{ulem}

\usepackage[final]{pdfpages}
\usepackage{authblk}
\date{}
\usepackage[sorting = none, style=numeric-comp]{biblatex}

\DeclareFieldFormat{title}{#1}
\DeclareFieldFormat{author}{#1}
\DeclareFieldFormat{journaltitle}{#1}
\DeclareFieldFormat{date}{#1}

\DeclareFieldFormat{booktitle}{#1}

\DeclareFieldFormat{url}{Available from\addcolon\space\url{#1}}

\DeclareFieldFormat{note}{#1}

\DeclareFieldFormat{url}{Available from\addcolon\space\url{#1}}
\DeclareFieldFormat{note}{#1}

\title{Evaluating the Relationship Between News Source Sharing and Political Beliefs}

\author{Sofía M del Pozo*$^{1,}$$^2$, Sebastián Pinto*$^{1,}$$^2$, Matteo Serafino$^3$,  Federico Moss $^1$, Tomás Cicchini $^{1,}$$^4$, Hernán A Makse$^3$ and Pablo Balenzuela$^{1,}$$^2$.
\par
{\normalsize $^*$ Equal contribution by these authors.}
\par
{\normalsize $^1$ Universidad de Buenos Aires, Facultad de Ciencias Exactas y Naturales, Departamento de Física. Buenos Aires, Argentina.}
\par
{\normalsize $^2$ CONICET - Universidad de Buenos Aires, Instituto de Física Interdisciplinaria y Aplicada (INFINA). Buenos Aires, Argentina.}

\par
{\normalsize $^3$ Levich Institute and Physics Department, City College of New York, 10031 New York, USA.
}
\par
{\normalsize $^4$ CONICET - Universidad de Buenos Aires, Facultad de Ciencias Exactas y Naturales, Instituto de Cálculo (IC)}}
\begin{document}
\maketitle

\begin{abstract}
In an era marked by an abundance of news sources, access to information significantly influences public opinion. Notably, the bias of news sources often serves as an indicator of individuals' political leanings. This study explores this hypothesis by examining the news sharing behavior of politically active social media users, whose political ideologies were identified in a previous study. Using correspondence analysis, we estimate the Media Sharing Index (MSI), a measure that captures bias in media outlets and user preferences within a hidden space. During Argentina’s 2019 election on Twitter, we observed a predictable pattern: center-right individuals predominantly shared media from center-right biased outlets. However, it is noteworthy that those with center-left inclinations displayed a more diverse media consumption, which is a significant finding. Despite a noticeable polarization based on political affiliation observed in a retweet network analysis, center-left users showed more diverse media sharing preferences, particularly concerning the MSI. Although these findings are specific to Argentina, the developed methodology can be applied in other countries to assess the correlation between users’ political leanings and the media they share.


\end{abstract}

\section{Introduction}
The massive and growing use of the internet and digital platforms has undoubtedly brought about changes in the way we live, including the resources at our disposal and the habits we incorporate \cite{graham2019society}. The political leaning of users play a key role in both sharing and accessing specific news items \cite{osmundsen2021partisan, eady2020news}. Indeed, the media bias observed in shared news on Twitter has been employed as a proxy for users' political ideologies \cite{cinelli2021echo, an2014sharing}.

News sharing behavior on Twitter within the context of political elections was analyzed by Weaver et. al.  \cite{weaver2019communities}. The authors used a bipartite network of users and news articles and analyzed the emergence of communities in the projection onto the news layer and identify their main features that explain these communities. A similar approach was addressed in \cite{cicchini2022news} for Argentinian media outlets, comparing electoral with non-electoral periods of time.  Within this framework, the researchers observed that groups of users on Twitter form based on their preferences for specific media outlets.

\par Inferring political leaning from available information of users in social media presents a challenge  that has been faced in different ways. For instance one approach involves utilizing hashtags information from tweets, as in \cite{cota2019quantifying}, to quantify the level  of support to the impeachment of the former Brazilian president. Another example is seen in \cite{zhou2021}, where hashtags were utilized to train a machine learning model aimed at predicting electoral trends during the 2019 Argentinian elections. Also, Cinelli et al., 2021 deduce users' political leanings based on the media bias of the news outlets they share on Twitter \cite{cinelli2021echo}. However, directly inferring user ideology from shared media bias remains a hypothesis awaiting validation. 

\par An alternative method for uncovering the political preferences of social media users was introduced by Barbera et al. (2015) \cite{barbera2015tweeting,barbera2015birds}, where they developed a Bayesian model. This model treats ideology as a latent variable, which can be inferred from observed connections among users, assuming that these connections adhere to the principle of homophily. Specifically, the authors estimated latent parameters by utilizing correspondence analysis on the adjacency matrix of users following political accounts on Twitter. A recent application of correspondence analysis is found in Flamino et al. (2023) \cite{flamino2023political}, where it was utilized to examine political polarization during the 2016 and 2020 US presidential elections. The authors estimated individual positions from the users-influencers adjacency matrix. Similarly, Falkenberg et al. (2022) employed a comparable method to estimate latent opinions within the online discourse surrounding measures to address climate change \cite{falkenberg2022growing}.

\par In this work, correspondence analysis is employed  not to infer users' ideology but to quantify their media preferences based on  the news articles they share. We compare these preferences with their political ideology previously inferred through a machine learning model developed and validated in Zhou et al. (2021) based only on the hashtags supporting one of the two coalitions that contested the elections that year \cite{zhou2021}. Specifically, we use users' connections to Argentinean news articles to deduce preferences for media outlets within a latent space, achieved by conducting correspondence analysis of the user-media matrix.  The comparison of both metrics allows us to measure how the political ideology of the users is related to the preference for the media they share. Additionally, we delve into the emergence of community structures within the retweet network (excluding retweets containing links to news articles) to discern whether user interactions are driven by political leaning or media outlet preferences.
This work aims to contribute to the quantitative analysis of the relationship between political ideology and media preferences, continuing the line of previous studies such as \cite{calvo2023winning} and \cite{aruguete2023news}.

\section{Background}\label{Bck}

\subsection{Argentinian Context}
\label{sec:argentina}

\par In this section, given our attention to Argentina, we delve into the country's political and media landscape during the 2019 presidential election campaign to provide essential contextualization.

\par Even though today Argentina is governed by Javier Milei of a libertarian party \cite{argentina_gov,mediabiasfactcheck_argentina}, over the last decade, Argentina's political landscape has been dominated by two major coalitions: one, a center-left coalition (CL) led by Cristina Fernández de Kirchner, known as \textit{Frente de Todos}, and the other, a center-right coalition (CR) led by Mauricio Macri, referred to as \textit{Juntos por el Cambio}. Cristina Kirchner held the presidency in Argentina during the periods of $2007-2011$ and $2011-2015$, while Mauricio Macri served as president from 2015 to 2019  \cite{murillo2020argentina,benezra2020neoliberalism,cantamutto2016}. During the 2019 elections, the center-left coalition presented Alberto Fernández and Cristina Fernández de Kirchner as their candidates. Meanwhile, the center-right coalition sought a second term for Mauricio Macri as president, with Miguel Ángel Pichetto as his vice-presidential candidate. \par National elections in Argentina comprise two obligatory phases: the primary election, known as PASO (which stands for \textit{Primarias, Abiertas, Simultáneas y Obligatorias} in Spanish, translating to Open, Simultaneous, and Obligatory Primaries in English), and the general election. In the year 2019, these events occurred on August 11th and October 27th, respectively. Additionally, if the results of the general election necessitate it, a third round, referred to as a \textit{ballotage}, may also be conducted.

\par Regarding the media landscape, the digital media scene in Argentina is primarily characterized by three major players: {\it Infobae}, {\it Clarín} and {\it La Nación}, each boasting approximately 20 million unique users in 2020, as reported by Comscore data \cite{comscore}. Following closely are a second tier of media outlets with audience numbers ranging from 6 to 13 million unique visitors. Prominent among this group are {\it Página 12}, {\it Ámbito Financiero}, {\it TN Noticias} and {\it El Destape Web}.

\par In Argentina, a pronounced polarization has been reported through the distinct  ideological orientations of the country's primary media outlets \cite{digitalnewsreport2023,cicchini2022news}. For instance, Página 12 is recognized as a left-of-center broadsheet newspaper \cite{bonner2018}, Infobae is considered as a center-left outlet \cite{mediabiasfactcheck_infobae}, while Clarín is considered a center-right tabloid \cite{mediabiasfactcheck_clarin}, and La Nación is characterized as a center-right broadsheet \cite{mediabiasfactcheck_lanacion,bonner2018}. 

\section{Data and Methods} 
In this section, we present the data description (section \ref{sec:Data}), which outlines the dataset utilized in our study, followed by the explanation about how we use this data and its study through correspondence analysis (section \ref{sec:methods}).

\subsection{Data description}
\label{sec:Data}

\par In this research, we employed a pre-existing  Twitter dataset \cite{zhou2021} comprising tweets collected between March 1, 2019, and August 27, 2019. 
The details of this dataset can be found in Appendix \ref{sec:original_dataset}.
From this dataset, we filtered for all types of tweets, including tweets, retweets, and quotes, that exclusively contained external URLs linking to Argentinian media outlets listed in the ABYZ News Links Guide \cite{abyz}. From this list we selected 17 media outlets by ranking the outlets in descending order by the number of times they were shared on Twitter and selecting those with broad recognition and influence not only on this platform but also across Argentine media as a whole, as reported in sources such as \cite{digitalnewsreport}. Given this filter, we first obtained a dataset encompassing the activity of 123,180 users, who collectively generated 1,039,281 tweets, sharing 66,982 unique news articles.
\par Secondly, we incorporated data concerning individuals' voting intentions, which had been previously determined using a model detailed in \cite{zhou2021}. In this paper, the authors developed a method to infer political preference of Twitter's users by implementing a dynamic classification model based on the balance of tweets in favor of each of the contending coalitions. Such a model, described in detail in \cite{zhou2021} provides a temporal label to a subset of 17,349 users  as supporters of the center-left (CL) candidates (Fernández-Fernández) and 15,361 individuals as sympathizers of the the center-right (CR) coalition (Macri-Pichetto). Supporters of the CL coalition shared 19,276 news articles, while those leaning towards the CR coalition shared 10,135.

\par
\begin{figure}[ht]
    \centering
    \includegraphics[width = \columnwidth]{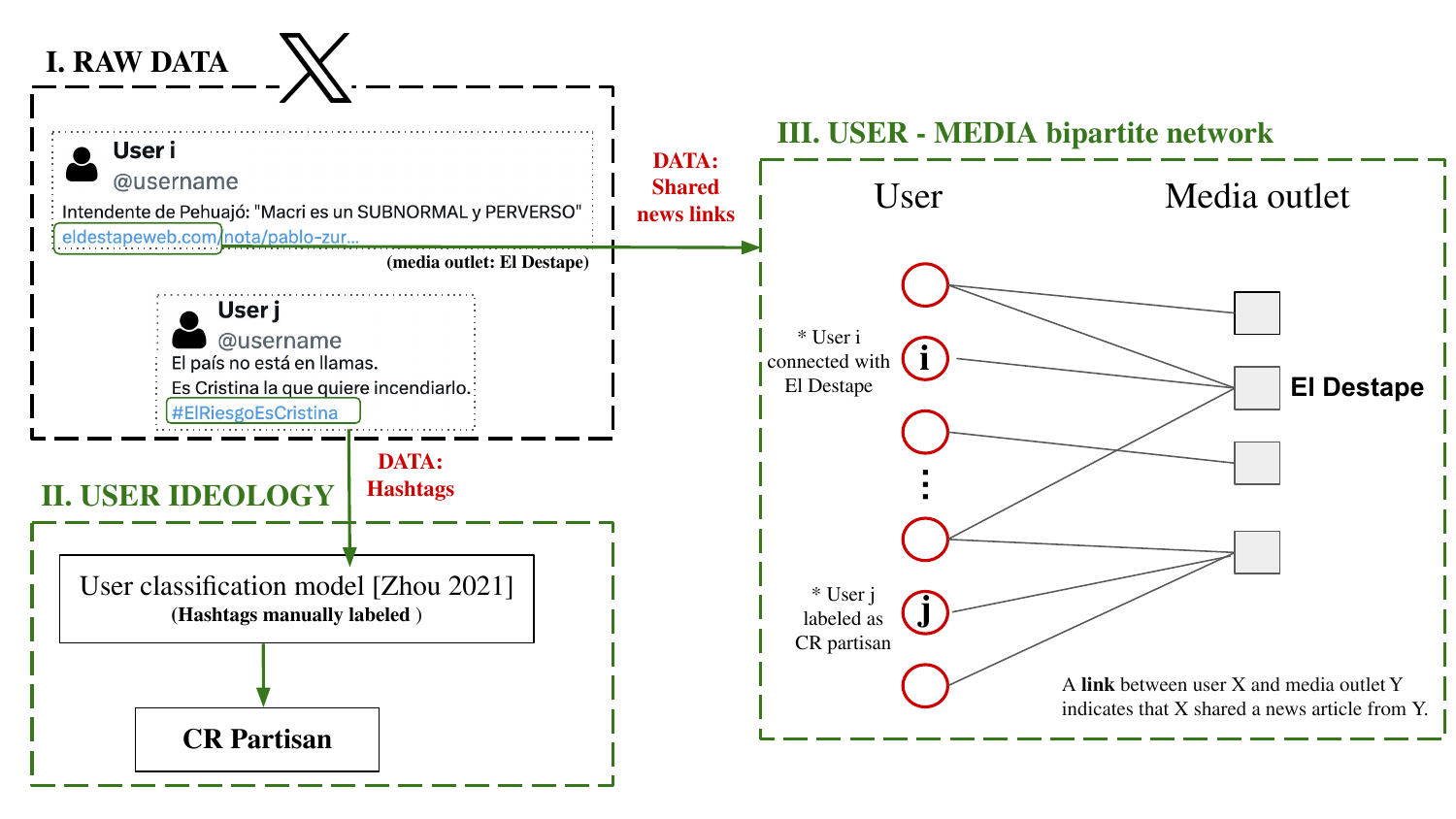}
    \caption{\textbf{Methodology pipeline.} \textbf{I. Raw data.} Example of original  data from Twitter (now X), with a  tweet sharing an URL to a news article at the top and another one with a political hashtag at the bottom. \textbf{II. User ideology.} Hashtags were used to train a logistic regression model to classify tweets as supportive of either candidate. Users are assigned to the candidate for whom they demonstrate the highest number of supportive tweets (further details in \cite{zhou2021}). \textbf{III. User-media.} The news URLs in the tweets are used to extract the media outlet. A bipartite network of users and media outlets is then created. For example, user {\bf i} is linked to the media outlet \textit{El Destape} because this user shared a news article from this media outlet.}
    \label{fig:method}
\end{figure}

 Figure \ref{fig:method} sketched the methodology followed to organize the large set of tweets sharing news of politically tagged users into a bipartite network of users-media outlets.

\subsection{Methods}
\label{sec:methods}

\par We organize the data in a user-media matrix, where each row is associated with a user and each column is associated with a media outlet. The components of the matrix represent the number of times a given user shares an article from a specific media outlet. By applying correspondence analysis (detailed below), we calculate what in this context we call the Media Sharing Index (MSI). This index positions users within a latent space reflecting their preferences for specific groups of media outlets. Simultaneously, it places media outlets within the same latent space, determined by the average preferences of their audience. Essentially, users closer in this space tend to share similar media outlets, indicating comparable preferences in media sharing. Similarly, media outlets situated closely in this latent space imply shared usage by a similar set of users. We compare the MSI with the political leaning of those users identified by \cite{zhou2021} and their position within the interaction network.

\subsubsection{Correspondence Analysis}
\label{sec:MSI}
\par

Following previous studies \cite{barbera2015tweeting, flamino2023political} that propose a methodology for inferring user coordinates in a latent space of social media based on correspondence analysis \cite{greenacre2010correspondence}, we begin by establishing a bipartite network denoted as G = (U, V, E), where U represents the set of users, V denotes the news outlets, and E stands for the edges in the graph. The corresponding adjacency matrix associated with this network is denoted as Y. The element $y_{ij}$ represents the number of times user $i$, with $i \in U$ shares a news from a media outlet $j$, with $j \in V $. 
The main difference between our implementation and the seminal proposal \cite{barbera2015tweeting}  is that the bipartite network here is 
based on the content of users' tweets, rather than explicit network connections (e.g., following-followers relations or retweets).

The matrix $Y$ is converted to the correspondence matrix $P$ by dividing each element by its grand total $P = Y/ \sum_{ij} y_{ij}$. The element $p_{ij}$ represents the probability of finding an event in which user $i$ shares an article from media outlet $j$. 
From matrix $P$, the matrix of standardized residuals $S$ is computed as:
\begin{equation*}
S = D^{1/2}_{r} (P - r c^{T}) D^{1/2}_{c}
\end{equation*}
where vectors $r$ and $c$ are defined such as $r_i = \sum_j p_{ij}$ and $c_j = \sum_i p_{ij}$.The element $r_i$ represents the likelihood that user $i$ shares an article from any media outlet. Conversely, $c_j$ represents the probability of media outlet $j$ being shared by any user.
The elements of outer product $rc^T$ ($r_ic_j$) can be interpreted as the probability of user $i$ sharing a media outlet $j$ given a null model where only the activity of user $i$ and the frequency of media outlet $j$ being shared matter.
By defining diagonal matrices $D_r = \text{diag}(r)$ and $D_c = \text{diag}(c)$, we can express the elements of $S$ as follows:
\begin{equation*}
    s_{ij} = \frac{p_{ij} - r_ic_j}{\sqrt{r_ic_j}}
\end{equation*}
This expression can be interpreted as the deviation, measured in standard units, of $p_{ij}$ from a null model where users and media outlets are independent. 

\par In order to compute the MSI for each user, we firstly perform singular value decomposition to $S$, that is
\begin{equation*}
    S = U D_\alpha V^T
\end{equation*}
where $UU^T = V^T V = I$ and $D_\alpha$ is a diagonal matrix with the singular values on its diagonal. The Media Sharing Index for the user i, $MSI_{i}$, is then identified by the standard row coordinates by projecting only over the first singular component:
\begin{equation*}\text{MSI}_{\text{i}} = (D_r^{1/2} U )_i
\end{equation*}
and finally normalizing these values to have zero mean and standard deviation equal to 1.  As such, users with similar values of MSI imply that they share a similar set of media outlets. 
In particular, if user sharing behavior is driven by two distinct groups of media outlets, as shown in \cite{cicchini2022news}, we would expect to observe a group of individuals with $\text{MSI}_i > 0$ and another with $\text{MSI}_i < 0$.

Finally, we define the MSI for media outlets as the weighted average of the MSI of the users, weighted by the number of times user $i$ shares media $j$:
\begin{equation}
\text{MSI}_j = \frac{\sum_i y_{ij} \text{MSI}_i}{\sum_i y_{ij}}
\end{equation}
The interpretation of \text{MSI}$_j$ is analogous to one provided for \text{MSI}$_i$: media outlets with $\text{MSI}_j > 0$ will have a very different set of users who share its articles compared to media outlets with $\text{MSI}_j < 0$.

\section{Results}

\subsection{Media Sharing Index}

\par As described above, we construct the bipartite adjacency matrix $Y$, where each element $y_{ij}$ represents the number of news articles from media outlet $j$ shared by user $i$. We apply correspondence analysis to this matrix to calculate the Media Sharing Index (MSI), as discussed in section \ref{sec:MSI}. For simplicity, we focus on the primary 12 media outlets, excluding media outlets that are shared by only a few users and where the majority of the shared articles come from the outlets themselves. This reduces the dataset to 59,874 unique news articles (approximately 88\% of the total unique news in the dataset) and 120,626 users (about 97\% of the users in the original dataset). These users originate from a total of 1,015,380 tweets containing links to one of these 12 main outlets, which constitutes about 98\% of the original tweet volume.

\par
\begin{figure}
    \centering
    \includegraphics[width = 0.9\columnwidth]{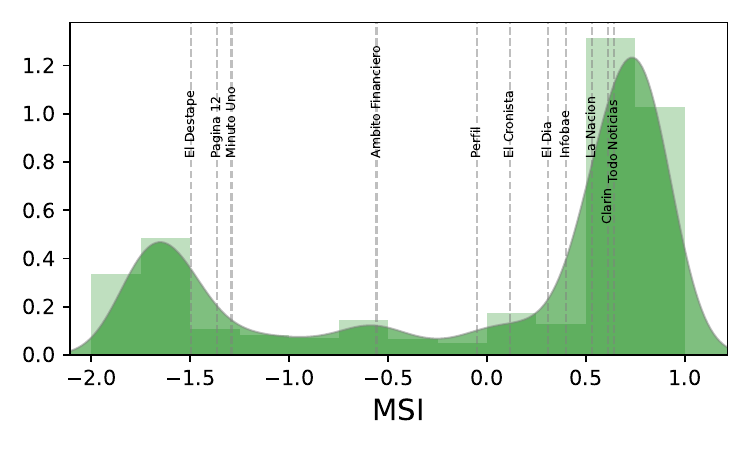}
   \caption{{\textbf{Probability density of the Media Sharing Index (MSI).} This graph shows the probability density of the MSI for users and the 12 main media outlets. Radio Mitre is excluded from this display due to its status as a positive MSI outlier. Histograms have been smoothed using a Gaussian kernel with a bandwidth of $0.15$. The grey lines indicate the positions of the media outlets.}}
    \label{fig:MSI}
\end{figure}

 Figure \ref{fig:MSI} illustrates the probability density function of the MSI for users who share articles from at least one of the primary 12 media outlets. This figure reveals the emergence of a bimodal distribution in the Media Sharing Index. Unimodality is rejected with a p-value practically equal to zero ($p<0.001$), as determined by the Dip test \cite{hartigan1985dip, diptest}.
This bimodal distribution reflects the preferences of users sharing content from two distinct groups of media outlets. Specifically, a majority of users share news from a group of outlets that includes Clarín, La Nación, Todo Noticias, among others, with an MSI close to $+1$. Conversely, a minority group prefers outlets such as El Destape, Página 12, and Minuto Uno, with an MSI close to $-1$.

\par Given that Clarín and La Nación are considered center-right outlets \cite{mediabiasfactcheck_clarin, mediabiasfactcheck_lanacion} and Página 12 is viewed as a left-of-center broadsheet \cite{bonner2018}, we can associate the Media Sharing Index with media bias along the left-right political dimension. These results prompt the question: do left-leaning users predominantly share news from left-leaning newspapers, aligning with their beliefs? Or is news sharing behavior independent of their political preferences? In other words, can the media bias reflected in the news users share serve as a proxy for their political ideology? This is the question we aim to answer in this paper.

\subsection{Media sharing and political preferences}
\label{sec:ideological_valence}

\par In this section, we explore the potential link between the sharing preferences of users, as observed in Fig. \ref{fig:MSI}, and their underlying political polarization. As described in section \ref{sec:Data}, the dataset used in this analysis is the same one employed by Zhou et al. (2021) to infer the political preferences of Twitter users based on their posts. The labels assigned to users in \cite{zhou2021} are dynamic, allowing for a more nuanced description of a user's ideology. 
Therefore, we define the ideology valence of a user $i$ ($IV_{i}$) as:
\begin{equation}
IV_{i} = \frac{\#CR_{i} - \#CL_{i}}{\#CR_{i} + \#CL_{i}}.
    \label{ideology_valence}
\end{equation}
Here, $\#CR_{i}$ and $\#CL_{i}$ denote the number of times user $i$, identified as a center-right or center-left partisan at that time, shared a news article, respectively. The sum $\#CR_{i} + \#CL_{i}$ represents the total number of news articles shared by user $i$. By this definition, $IV_{i} = 1$ indicates a user consistently labeled as center-right, representing a pure CR partisan. Conversely, $IV_{i} = -1$ indicates a pure CL partisan. $IV$ is only defined for users with a label assigned in \cite{zhou2021}.

\par
\begin{figure}
  \centering
    \includegraphics[width = 0.7\columnwidth]{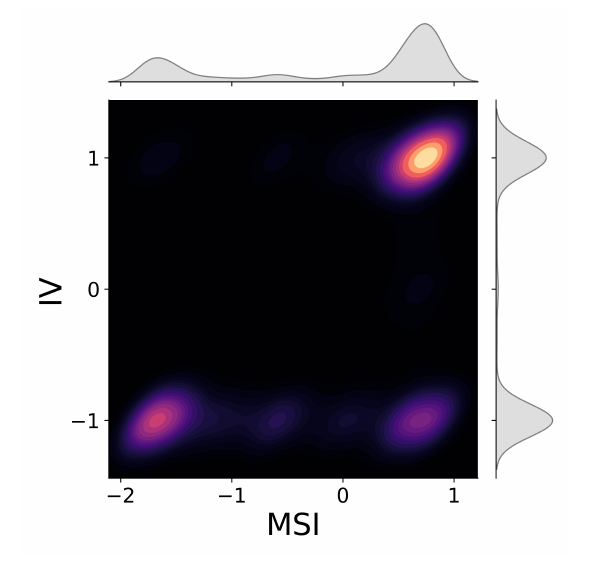}
\caption{\textbf{Joint Probability Density of MSI and IV.} This figure illustrates the relationship between the media sharing behavior of users and their ideological leaning.}
  \label{fig:Ideology}
\end{figure}

 The relationship between the ideology valence (IV) and the Media Sharing Index (MSI) is illustrated in Fig. \ref{fig:Ideology}. As previously mentioned, the MSI can serve as an indicator of media bias along the left-right political dimension. However, what Fig. \ref{fig:Ideology} reveals is that the media sharing behavior of users on social media cannot necessarily be directly associated with their ideological leanings.
This figure demonstrates that while center-right users exhibit a distinct media sharing profile, with a preference for sharing outlets aligned with their political leanings, center-left users share news from media sources corresponding to both peaks observed in the probability density of the MSI in Fig. \ref{fig:MSI}.
 
\par A possible interpretation of the asymmetric behavior observed between ideological groups is that media outlets with MSI $> 0$, such as Clarín, La Nación, and Infobae, are also predominant players in the Argentine media landscape \cite{comscore} (see section \ref{sec:argentina}). These outlets’ extensive reach and influence may contribute to the observed sharing patterns among center-left users at the aggregate level.

\subsection{Retweet user networks}

\par In this final section, we analyze the interaction among users by constructing the retweet network. We specifically focus on the relationship between the emerging community structure, the Media Sharing Index (MSI), and the users' political preferences.
\par By definition, the retweet network is directed and weighted. The direction of the links indicates the flow of information (i.e., arrows point from a retweeted user to the user who retweets), and the weights reflect the number of retweets between users. This network comprises 114,673 nodes, representing approximately 90\% of the users described in section \ref{sec:Data}. The remaining 10\% are users who did not retweet or were not retweeted by any other users during the analyzed period. Additionally, there are 12,993,644 edges, corresponding to the total number of retweets among users in this network. It is important to note that this network was constructed only considering retweets that do not include links to news articles, meaning it contains no data used in the calculation of the MSI.

\par
\begin{figure}
    \centering
    \includegraphics[width = \columnwidth]{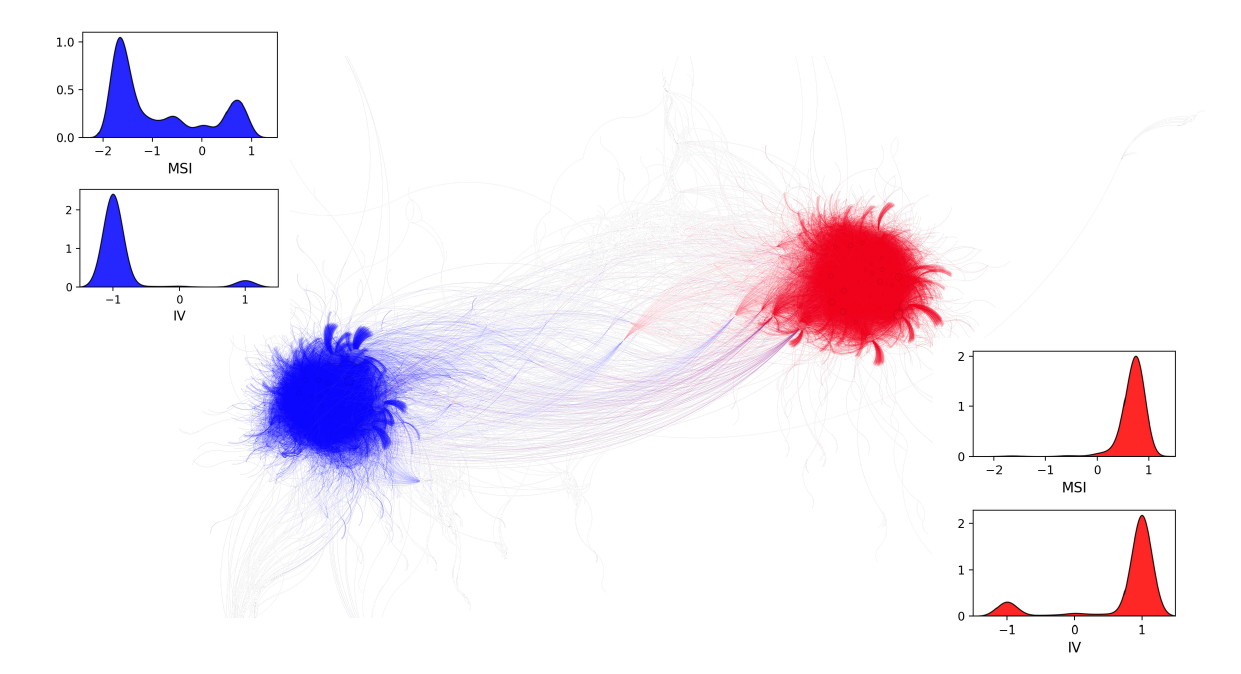}
\caption{\textbf{Retweet network.} The two largest communities detected by the Louvain algorithm are displayed (account for 75\% of the entire network). Histograms illustrate the distribution of the MSI and IV for each community. Based on these distributions, the red community can be associated with a center-right political leaning, while the blue community can be associated with a center-left leaning.}
    \label{fig:retweet_network}
\end{figure}

 In Fig.  \ref{fig:retweet_network}, the two main communities within the retweet network are shown, identified using the Louvain algorithm \cite{blondel2008fast}. Although the algorithm detects 440 communities in total, the two largest communities account for 75\% of the entire network, with nearly an equal number of nodes in each community.
Figure \ref{fig:retweet_network} also reveals a highly modular structure of the retweet network, with a modularity score of approximately $Q \sim 0.48$.

\par Histograms in Fig. \ref{fig:retweet_network} reveal the profile of each community in relation to the Media Sharing Index and the Ideological Valence. Each community correlates strongly with a distinct ideological position: The red community in Fig. \ref{fig:retweet_network}, representing 38$\%$ of the network, aligns with the center-right,  while the blue community, representing 37$\%$ of the network, aligns with the center-left. 
The fact that the histograms for the Ideological Valence index show a clear peak for both communities suggests that the community structure emerging from the retweet network is a reliable proxy for the ideological positions of its members.

\par The association found between the community label and the ideological position of users supports the reproduction of the results discussed in section \ref{sec:ideological_valence} at a community level: the center-right community exhibits media sharing behavior favoring center-right media outlets, as depicted in the blue histogram of the MSI. Conversely, the center-left community displays a less biased MSI distribution, indicating that this community shares content from both center-left media outlets and those biased towards the opposite ideological spectrum. 
As mentioned in section \ref{sec:ideological_valence}, the observed diversity in sharing patterns among the CL group may be linked to the prominence of CR media outlets in the overall media ecosystem.

\section{Discussion and Conclusions}

\par In this work, we describe the collective news sharing behavior of thousands of Twitter's users by a coordinate in a latent space which we named Media Sharing Index (MSI). This is obtained by performing correspondence analysis to the bipartite network of users - media outlets which emerge from the information of news articles shared by users.

\par The MSI metric enables us to set up a scale that describes the preferences of social media users in news sharing behavior. In this work, we specifically observe a bimodal distribution of the MSI, where the two clearly defined peaks can readily be linked to two distinct groups of media outlets. In the case of Argentina, these two groups are exemplified by Clarín, La Nación, Infobae, and Todo Noticias, among others, on one side, and on the other side, Página 12 and El Destape. These six media outlets stand out as the most widely shared on social media. Given that  Clarín and La Nacion are considered as center-right outlets \cite{mediabiasfactcheck_clarin, mediabiasfactcheck_lanacion} and Página 12 as a left-of-center broadsheet newspaper \cite{bonner2018}, we can associate the Media Sharing Index as a measure of media bias in the left-right political dimension.

\par The strength of the Media Sharing Index lies in its ability to quantify news sharing behavior among diverse media outlet groups. Coupled with the inference of users' political leaning derived from the machine learning model outlined in \cite{zhou2021}, we can explore the correlations between these dimensions and address questions such as: \textit{Can the bias of the shared media outlets be used as a proxy of user's ideology?} 

\par Our analysis reveals distinct patterns: while the CR group predominantly shares news from media that align with their ideological stance, the CL group exhibits a broader range of media sources in their sharing behavior. This observed diversity could likely be influenced by the prominent role of CR media outlets within the broader media ecosystem in Argentina. Accordingly, this ideological asymmetry in news preferences and exposure has been also widely documented in other contexts. For instance, in the US, studies have examined the media ecosystem \cite{benkler2018network, jamieson2008echo} and the preference for news exposure in social media. In \cite{hmielowski2020asymmetry}, the authors analyze the reinforcing effects of liberal and conservative media on political beliefs during the 2016 US election, finding that conservative beliefs contribute more to a conservative media echo chamber than liberal beliefs do to a liberal one. Similarly, \cite{gonzalez2023asymmetric} explores ideological segregation in political news on Facebook during the 2020 US election, revealing that conservatives consume a disproportionately large amount of news, including most misinformation, while liberal sources are less prevalent. Additionally, findings regarding the sharing of misinformation and fake news indicate that, during the 2016 US election, conservative Facebook users were more likely to share pro-Trump fake news than their liberal or moderate counterparts \cite{guess2019less}. Research conducted in countries outside the US also suggests that ideology has a stronger influence on right-wing users compared to left-wing users when it comes to news-sharing behavior \cite{aruguete2019news}.

\par The most significant limitation of this study is that it is restricted to Argentina during the 2019 electoral period. It is essential to examine how these results vary in other electoral periods in Argentina and what the results would be in different countries. Nonetheless, we have developed a methodology here that can be extrapolated to all these scenarios in future studies.

\section*{Funding}

HAM and MS were supported by NSF Grant No. 2214217. PB, SMdP and SP were supported by PICT-2020-SERIEA-00966. 

\section*{Author contributions statement}

SMdP, SP, and MS were responsible for collecting the raw data. SMdP and SP developed the computational code utilized consistently throughout the paper.
TC and FM made contributions to the statistical analysis. 
PB and HAM conceptualized the research. All authors engaged in discussions about the results and collaborated on the development of the manuscript.

\section*{Data and code availability statement}

\par Twitter data analyzed in this work are provided according to its terms and are available at https://osf.io/u29tk/.
Analytical codes are available in https://github.com/spinto88/MediaSharingIndex.

\appendix
\section{Original dataset}
\label{sec:original_dataset}

\par As mentioned in \ref{sec:Data}, we employed a pre-existing Twitter dataset \cite{zhou2021} comprising tweets collected between March 1, 2019, and August 27, 2019. This dataset was obtained through the Twitter Academic API (at the time of acquiring the data, this API allowed full access to the complete history of the social media platform) by searching for the following queries: 
\textit{Pichetto, Espert, victoria, Bregman, argentina, Voto, Kirchner, Propuesta Republicana, Moreno, Peron, Sola, kirchnerismo, Massa, UCR, PASO, Alternativa Federal, Justicialista, PRO, Movimiento Socialismo, Rossi, \#ArgentinaVota, Union Civica Radical, fuerza, Votar, Vidal, Alfonsin, alferdez, kicillof, Scioli, vamos, Frente Izquierda, Donda, miguelpichetto, CFK, Vote, kicillofok, Alberto Fernandez, Urtubey, elecciones, sergiomassa, ganamos, Peronista, CFKArgentina, peronismo, \#EleccionesArgentina, PJ, Lavagna, Partido Justicialista, mauriciomacri, Consenso 19, Cristina, Lousteau, Cambiemos, macrismo, mariuvidal, Capitanich, apoyo, Macri}.
Exclusively, tweets written in Spanish were selected.

\par The full dataset analyzed comprised around 170 million tweets and about 3.5 million users. Although the dataset is massive, given the keywords used to obtain it, the dataset may include users with a certain bias towards tweeting about politics and, therefore, must be considered with caution when evaluating its representativeness of the Argentinian population. On the other hand, Zhou et al. \cite{zhou2021} reported an average classification per day of 114,653 tweets as CR supporters and 96,576 as CL supporters (MP and FF, respectively, in \cite{zhou2021}), which indicates a balanced classification.

\par All of these numbers are drastically reduced when considering only those tweets with URLs linking to Argentinian outlets, as stated in section \ref{sec:Data}.

\printbibliography

\end{document}